\documentclass[11pt,a4paper]{article}
\usepackage{jheppub}
\usepackage{amsmath}
\usepackage[most]{tcolorbox}
\usepackage{dsfont}
\usepackage{ulem}
\usepackage{natbib}
\usepackage{xcolor}
\usepackage[hang,flushmargin]{footmisc}
\usepackage{tikz-cd}
\usepackage{enumitem}
\usepackage{amsfonts,amssymb, amscd,amsmath,latexsym,amsbsy,bm}
\usepackage{stmaryrd}
\usepackage{todonotes}
\usepackage{float}

\makeatletter\renewcommand{\@biblabel}[1]{#1.}\makeatother

\newtcolorbox{empheqboxed}{colback=gray!20, 
 colframe=white,
 width=\textwidth,
 sharpish corners,
 top=0mm, 
 bottom=0pt
}

\title{Liouville integrable binomial Hamiltonian system }

\author{   Mustafa Mullahasanoglu}
\affiliation{
Department of Physics, Bogazici University,\\ 34342 Bebek, Istanbul, Turkey\\[-0.4cm]}
\emailAdd{mustafa.mullahasanoglu@boun.edu.tr}

\abstract{In this study we work on a novel Hamiltonian system which is Liouville integrable. In the integrable Hamiltonian model, conserved currents can be represented as Binomial polynomials in which each order corresponds to the integral of motion of the system. From a mathematical point of view, the equations of motion can be written as integrable second-order nonlinear partial differential equations in $1+1$ dimensions.}

\begin{document}
\maketitle
\flushbottom

\section{Introduction}
Recently, in \cite{Gahramanov:2017att}, there appears a connection between integrals of motions of the integrable Hamiltonian system and Motzkin polynomials in which coefficients of the polynomials can be depicted as a path on a unit lattice. Each pattern on the lattice and corresponding polynomial is introduced as the integral of motion. The polynomial property \cite{Gahramanov:2017att, motzin} of the Motzkin numbers\footnote{Motzkin numbers and its path and polynomial properties are studied in different fields, e.g. in the mathematics \cite{krat, flajolet} and in physics \cite{blythe, bravyi, francesco2010q}.} (the sequence A001006 in \cite{OEIS})  $1, 2, 4, 9, 21, 51 ...$ is the following 
\begin{equation}
    P_{n}=\sum_{k=1}^{n} \alpha^{k}\beta^{n-2k}t_{n,k}\:,
\end{equation}
where 
\begin{equation}
 t_{n,k}=\frac{(n-2)!}{(n-2k)!k!(k-1)!}    
\end{equation}
Some of the polynomials are
\begin{align}
    P_{2}&=\alpha ~~~~
    P_{3}=\alpha\beta \quad 
    P_{4}=\alpha\beta^2+\alpha^2 \quad
    \nonumber \\
    P_{5}&=\alpha\beta^3+3\alpha^2\beta \quad
    P_{6}=\alpha\beta^4+6\alpha^2\beta^2+2\alpha^3
\end{align}
where one can pick $P_{1}=\beta$. 

If one re-parametrize\footnote{If one leaves as in $\pi$ and $\phi'$  in polynomials, coefficients are Narayana numbers \cite{euleriannumbers} (the sequence A001263 in \cite{OEIS}) which are also related to Motzkin numbers.} as $\pi \phi'=\alpha$ and $\pi+\phi'=\beta$ where $\pi=\pi(x,t)$, $\phi=\phi(x,t)$ and $\phi'$ is the derivative of $\phi$ with respect to $x$, the integrals of motions can be written in the form
\begin{equation}
    I_{n}=\int P_n dx\:,
\end{equation}
 where the Hamiltonian of the system is
\begin{equation}
    H=\int (\pi^2 \phi' +\pi \phi'^2 )dx\:,
    \label{motzkinhamiltonian}
\end{equation}
The corresponding equations of motion are
\begin{equation}
    \dot{\pi}=2\pi\pi'+2\pi'\phi' +2\pi\phi'' \:,\quad
    \dot{\phi'}=2\phi'\phi''+2\pi'\phi' +2\pi\phi''\:,
    \label{motzkineom}
\end{equation}
where the dot represents the time derivative. 

By introducing new variables  $u=\pi$ and $v=\phi'$ one obtains symmetric coupled differential equations (\ref{motzkineom})
\begin{equation}
    u_{t}=(u^2+2uv)_{x}\:, \quad
    v_{t}=(v^2+2uv)_{x}\:,
\end{equation}
where indices represent derivatives. Note that similar equations were studied using a hydrodynamic perspective  for the Born-Infeld equations in \cite{arik:1989, olver, Sheftel1986, dubrovin}.

The Hamiltonian (\ref{motzkinhamiltonian}) under the limit $\pi>>\phi$ gives a Hamiltonian \cite{Akhmedov:2010mz} (see also, e.g. \cite{Akhmedov:2010sw}) for the simplest case, matrix scalar field theory, which is written to describe RG flow equations. In a field theory under change of scale \cite{Polchinski:1983gv, Bervillier:2004mf}, the exact renormalization group studied by the Polchinski equation allows the investigation of dynamics of operators. The equations coming from this reduced Hamiltonian \cite{Akhmedov:2010mz} describe shock waves and governing equations are known integrable. 

In this paper, we investigate integrability properties motivated by the above discussion of binomial numbers by starting the definition of Hamiltonian

\begin{equation}
\begin{aligned}
    H=\int (c\pi\phi'+a\pi+b\phi')^2dx 
\end{aligned}
\end{equation}
where $a$, $b$, and $c$ are free parameters. Integrals of motions for such system have the following form
\begin{equation}
\begin{aligned}
    I_n=\int (\alpha+\beta)^ndx \:, 
\end{aligned}\label{iombinomial}
\end{equation}
where $I_2=H$ and free variables are fixed to $1$ in this notation. However, it is important to remark that one can have integrable models such as $I_n=\int \alpha^ndx$, and removing free variables by simple redefinition leads to the loss of the general model.   

These integrals of motions (\ref{iombinomial}) corresponding infinite degrees of freedom are in involution as required for Liouville integrability, i.e. any two integrals of motion $I_n$ and $I_m$ commute with respect to Poisson brackets defined for the field theory. 



Briefly, we observe that one can use binomial numbers to construct an integrable Hamiltonian system like the Motzkin numbers in which its integrability is discussed in \cite{Gahramanov:2017att}. The main result is that the binomial Hamiltonian system is completely Liouville integrable. Possible further studies for the novel integrable model are presented in the conclusion section. 

\section{Binomial integrable model}

Liouville integrability has two basic conditions that must hold simultaneously for a dynamical system \cite{Arnold,torrielli2016lectures,hoppe}
\begin{itemize}
    \item number of degrees of freedom = number of integrals of motion $I_{n}$\:.
    \item $\{I_{n},I_{m}\}=0$ that means integrals of motion are in involution.
\end{itemize}
In a field theory, the Liouville integrable model needs to have an infinite number of integrals of motion since there are infinite degrees of freedom.
If the existence of a sufficient  number of integrals of motion is satisfied, the integrability property depends on whether the integrals of motions are in involution in which arbitrary two integrals of motions of the system in the Poisson bracket vanish.

Here, we introduce the binomial Hamilton system with the general formulation of the integrals of motions


\begin{equation}
I_{n}=\int (c\pi\phi'+a\pi+b\phi')^ndx \:,
\end{equation}
where $a$, $b$, and $c$ are free parameters and the Hamiltonian is
 \begin{equation}
H=I_{2}=\int (c\pi\phi'+a\pi+b\phi')^2dx \:,
\end{equation}
where equations of motion are 
\begin{equation}
\begin{aligned}
    \dot{\pi}=2[(c\pi+b) (c\pi\phi'+a\pi+b\phi')]'\:, \\
    \dot{\phi'}=2[(c\phi'+a) (c\pi\phi'+a\pi+b\phi')]'\:,
\end{aligned}
\end{equation}
where the dot and prime represent time and spatial derivatives, respectively.


Integrable equations of motion with reparametrization $u=\pi$ and $v=\phi'$ can be written as

\begin{equation}
    u_{t}=2[(cuv+au+bv)(cu+b)]_{x} \:,\quad
    v_{t}=2[(cuv+au+bv)(cv+a)]_{x}\:.
    \label{eomb}
\end{equation}

The simplest form of these equations can be written  by setting $a,b,c=1$
\begin{equation}
    u_{t}=2[u^2v+u^2+2uv+u+v]_{x} \:,\quad
    v_{t}=2[v^2u+v^2+2uv+u+v]_{x}\:.
\end{equation}

First, we check that we have integrals of motion satisfy, i.e. $\dot{I}_{n}=0$, 
\begin{equation}
\begin{aligned}
    \dot{I}_{n}= \int \frac{d}{dx}  \left[\frac{6nc}{n+1}(c\pi\phi'+a\pi+b\phi')^{n+1}+4ab(c\pi\phi'+a\pi+b\phi')^{n} \right] dx=0\:.
    \end{aligned}
    \end{equation}

Then the definition of the Poisson bracket \cite{hoppe} 
  \begin{equation}
\{I_{n}, I_{m}\}=\int  \left( \frac{\delta I_{n}}{\delta \pi} \frac{\delta I_{m}}{\delta \phi}-\frac{\delta I_{m}}{\delta \pi} \frac{\delta I_{n}}{\delta \phi} \right) dx\:.
\end{equation}
used to determine whether the second condition of Liouville integrable models holds.

Due to the definition of the integrals of motions, the calculation of variational derivative can be written in the short form
\begin{equation}
\{I_{n}, I_{m}\}=\int  \frac{\partial I_{n}}{\partial \pi } \frac{d}{dx}\left(\frac{\partial I_{m}}{\partial \phi'}\right)-\frac{\partial I_{m}}{\partial \pi} \frac{d}{dx}\left(\frac{\partial I_{n}}{\partial \phi'}\right)dx \:.
\end{equation}
The calculations reveal that

\begin{equation}
\begin{aligned}
\{I_{n}, I_{m}\}=\int \frac{d}{dx}  \Bigg[&\frac{(m-n)mnc}{m+n-1}(c\pi\phi'+a\pi+b\phi')^{m+n-1} \\
&+\frac{(m-n)mnab}{(m+n-2)}(c\pi\phi' +a\pi+ b\phi')^{(m+n-2)} \Bigg]dx
\\&=0\:.
\end{aligned}
\end{equation}

We showed that the integrals of motion are in involution. Therefore one can conclude that the binomial Hamiltonian model is completely Liouville integrable.

\section{Conclusion}

The idea of the correspondence between sequences such as Motzkin and the integrable Hamiltonian model allows us to construct a new integrable model by using the binomial polynomials for the integrals of motion. The equations of motion of the model are also new integrable differential equations \cite{shabat}.

A similar approach can be used for the Dyck or the Riordan paths and corresponding sequences to
construct integrable systems. On the other hand, one can use symmetric polynomials, or the models introduced in this paper can be generalized to the theory with more scalar fields $\phi_i$ (or in higher dimensions).

As a further study, both Motzkin and Binomial integrable systems are studied via Liouville criteria and it would be interesting to construct Lax pairs \cite{hoppe} for these systems\footnote{One can search for the Lax pairs by the use of machine learning techniques \cite{Krippendorf_2021}.}.


\section*{Acknowledgments}
We thank Ilmar Gahramanov for suggesting the problem, and for his many insightful teachings. We also thank Betül Gürbüz and Merve Yılmaz for their help and contributions to the ideas presented here. We are also grateful to George Jorjadze, Edvard T. Musaev, Mansur I. Ismailov, and O. Teoman Turgut for their valuable discussions.  We would like to thank Hesam Soltanpanahi for the warm hospitality at the Jagiellonian University (Krakow, Poland), where part of this work was carried out. Mustafa Mullahasanoglu is supported by the 2209-TUBITAK National/International Research Projects Fellowship Programme for Undergraduate Students under grant number 1919B011902237.

\appendix

\bibliographystyle{utphys}
\bibliography{bibtex}



\end{document}